# Modeling and computer simulation of the mixing and heat transfer in heterogeneous turbulent two-phase jets of mutually immiscible liquids by the method of Professor Alfred I. Nakorchevskii. Part 1

Ivan V. Kazachkov[1,2]
[1]Educational Research Institute of Natural Sciences and Economics,
Nizhyn Gogol State University, Ukraine, kazachkov.iv@ndu.edu.ua, https://en.ndu.edu.ua/
[2]Dept of Energy Technology, Royal Institute of Technology, Stockholm, Sweden,
ivan.kazachkov@energy.kth.se, http://www.kth.se/itm/inst?l=en_UK

SUMMARY

The present paper is devoted to the mixing and heat transfer features of mutually immiscible liquids in the two-fluid turbulent heterogeneous jet flow. Many natural and technical processes deal with the turbulent jets of mutually immiscible liquids, which represent an important class of the modern multiphase system dynamics. Differential equations for the axially symmetrical two-dimensional stationary flow and the integral correlations in a cylindrical coordinate system were considered for the jet from a nozzle into a space filled with another fluid that is not miscible with the first one. Parameters of the turbulent mixing in the two-phase jet flow were modeled and analyzed. The results may be of interest for some research and industrial tasks, where the calculation of parameters of the multiphase turbulent mixing and heat transfer are important.

## 1. STATEMENT OF THE MODELING OF MULTIPHASE TURBULENT FLOW

All parameters $a^l(t)$ (density of liquid, flow velocity, temperature, etc.) of a mixture in the turbulent multiphase flow are considered in accordance with the method proposed by Prof. Nakorchevski [1], developed and reported in many publications, e.g. [2-7]. The analog of the Navier-Stokes equations in a boundary layer approach was derived in the following form [1]:

$$\sum_{i=1}^{m}\left[\frac{\partial}{\partial x}(y\rho_i B_i u_i)+\frac{\partial}{\partial y}(y\rho_i B_i v_i)\right]=0, \qquad \sum_{i=1}^{m}\rho_i B_i\left(u_i\frac{\partial u_i}{\partial x}+v_i\frac{\partial u_i}{\partial y}\right)=-\frac{dp}{dx}+\frac{1}{y}\frac{\partial}{\partial y}\left[y\sum_{i=1}^{m}B_i\tau_i\right]_m , \tag{1}$$

where $a^l(t)=\sum_{i=1}^{m}B_i(t)a_i^l(t)$, $\sum_{i=1}^{m}B_i=1$. The function-indicator $B_i(t)$ was introduced for the phases in multiphase flow as follows:

$$B_i(t)=\begin{cases}1, & \text{if } i-\text{phase occupies the elementary volume } \delta\text{V}\\ 0, & \text{if } i-\text{phase is outside the elementary volume } \delta\text{V}\end{cases}.$$

In the stationary equations (1) for the flow of incompressible liquids, written in a cylindrical coordinate system, are: $p$- pressure, $\rho$- density, $u$,$v$- the longitudinal and transversal velocity components, $\tau_i$ - turbulent stress for the $i$-phase. All parameters of the flow are averaged on the characteristic interval by time chosen. Index $m$ belongs to the values at the axis of the flow (symmetry axis). The function-indicator of a phase in multiphase flow may be considered as the mathematical expectation, in contrast to the other multiphase approaches [8-10], which are based on the introduction of the volumetric specific content of a phase in multiphase flow. Nevertheless, use of the function-indicator allows computing the volumetric specific content of the phases, which have been introduced by another multiphase approaches.

Development of the model for two-phase jet of immiscible liquids is done according to the Fig. 1 and schematic representation in Fig. 2, where $r_0$ – radius of a nozzle, $u_{01}$- velocity at the nozzle. The conical surface 1 in Fig. 1 is a boundary of a homogeneous potential core $a$, the internal sublayer $b_1$ contains an ejected liquid as a disperse phase, while the external sublayer $b_2$ contains the liquid outgoing from the nozzle as a disperse phase. The internal and external sublayers are divided by the surface of phase inverse 2; and the surface 3 is dividing the turbulent and laminar flow zones 3, which is the most indefinite one; 4 is an external conical surface of the axisymmetrical mixing zone (conditionally smooth). Due to a limited regularity of the processes occurring in turbulent jets, the surfaces 2 and 3 in Fig. 1 are blurred into the corresponding regions of inversion and intermittency. The external boundary of the jet is the outer envelope surface 4 of the set of surfaces 3. The uniform velocity profile is assumed for the first liquid going from the nozzle. The surrounding liquid (phase 2) is in the rest before the first liquid starts flowing from the nozzle.

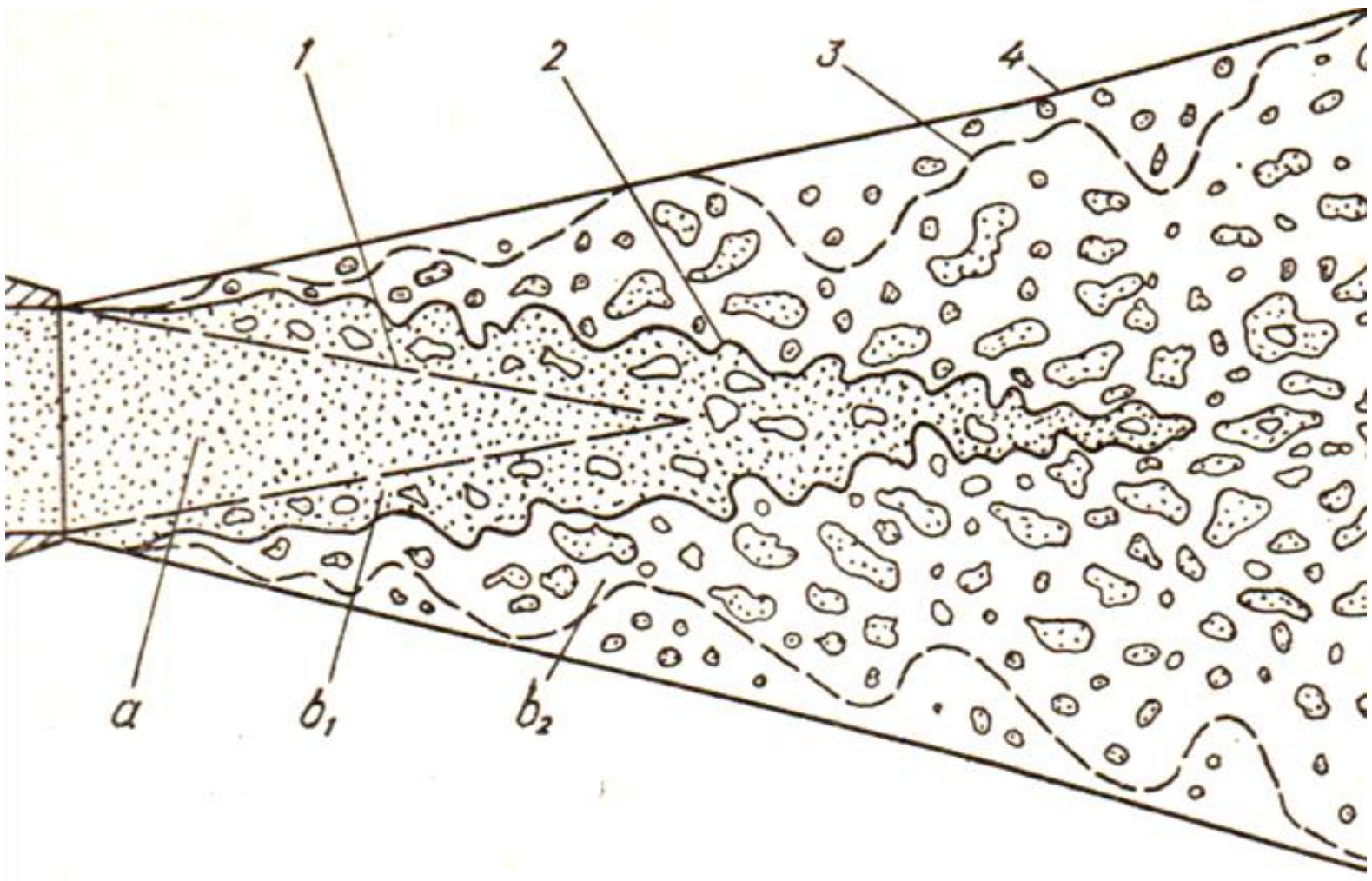

Fig. 1 General view of the multiphase turbulent jet in the pool of other liquid

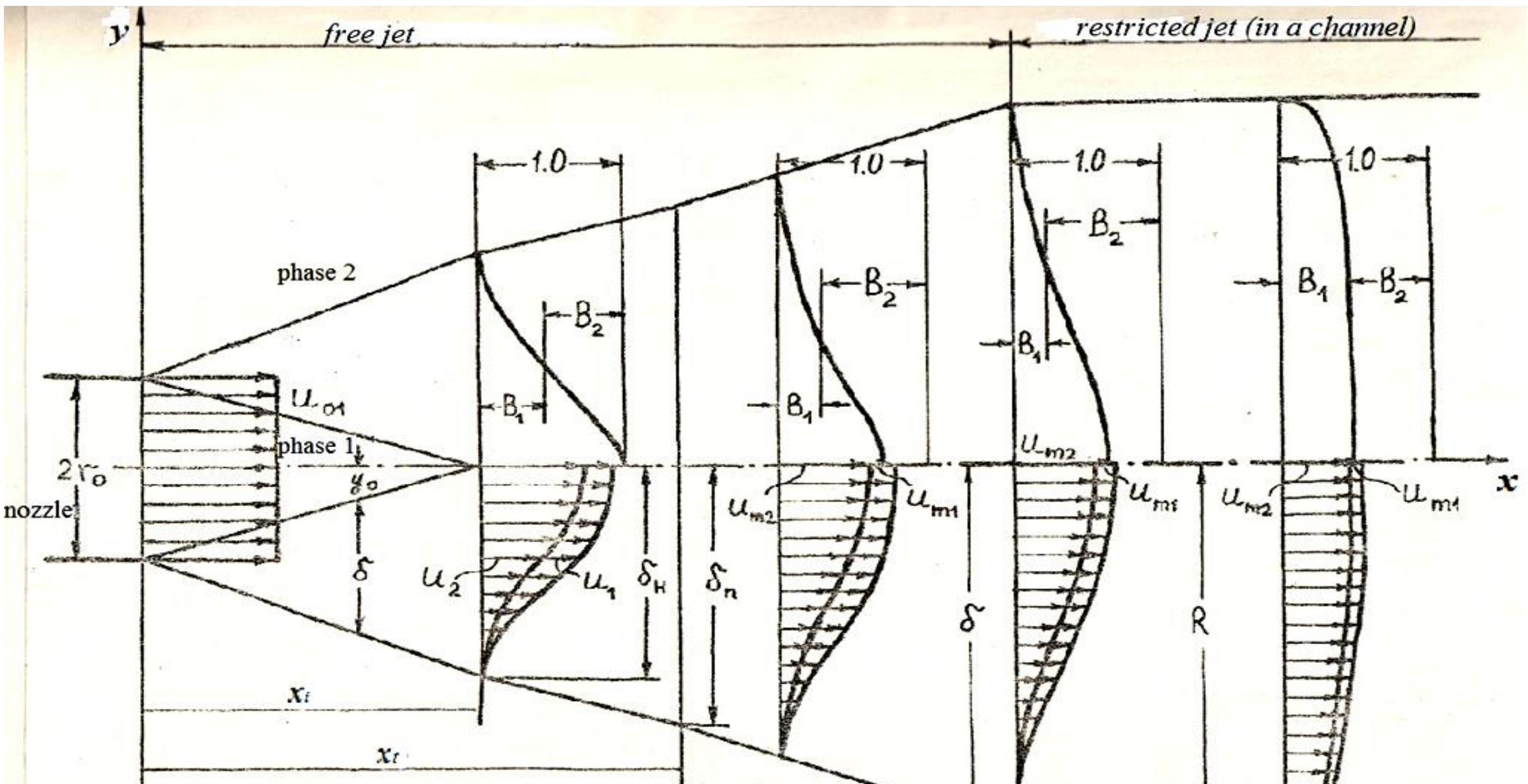

Fig. 2 Schematic representation of multiphase turbulent jet confined by channel at the distance $x_g$

The structural scheme for the mixing process in Fig. 2 is simplified: the initial part of the length $x_i$ with the approximately linear boundaries for the conical surface (in cylindrical coordinate system) of the internal core of a first phase and mixing zone between internal and external boundaries of the jet. The turbulent zone contains fragments of the phases as far as immiscible liquids have behaviors like the separate phases, with their interfacial multiple surfaces. The first phase in a potential core is totally spent in an initial part of the mixing zone. Then a short transit area follows. Afterwards the ground part of the two-phase jet begins, with the two phases well mixed across the entire layer of a jet.

Except for the parameters of phases, the function-indicator of phase $B_i(t)$ which shows the influence of *i*-th phase at each point of a space. Normally the spatial averaging of the conservation equations of mass, momentum and energy is performed for a description of multiphase flows based on the concept of volumetric phase content [8-10], which does not fit so well to the experimental study of a movement of the separate phases in a mixture. In contrast to this, an approach [1] with its special experimental technology and a micro sensor for the measurements in two-phase flows fits well for such flows. Actually all known methods of multiphase flows are well connected and the parameters averaged by time [1] can be easily transformed to the ones averaged by space [8-10].

The external boundary of a mixing zone is determined by zero longitudinal velocity of the second phase and zero transversal velocity of the first phase (the second phase is sucked from an immovable surrounding into a mixing zone). The function-indicator of the first phase $B_1(t)$ is zero at the external interface because it is absent in surrounding medium. Similar, the function-indicator $B_2(t)$ is zero on the boundary of the potential core, the interface of the first phase flowing from the nozzle. In a first approach, an influence of the mass, viscous and capillary forces are neglected.

With account of the above-mentioned, the boundary conditions are [1]:

$$y=y_0,\ u_i=u_{0i},\ v_i=0,\ \tau_i=0,\ B_1=1,\ \partial B_1/\partial\eta=0;\quad y=y_0+\delta,\ u_i=0,\ v_i=0,\ \tau_i=0,\ B_1=0. \qquad (2)$$

And dependence of the function-indicator $B_1$ from the longitudinal coordinate $x$ is introduced through the second derivative of it at the boundary of a jet $y=y_0$: $\partial^2 B_1/\partial\eta^2=h(x)$.

## 2. APPROXIMATION OF VELOCITY PROFILES AND FUNCTION-INDICATOR

The turbulent stress in the phase is stated by the "new" Prandtl's formula $\tau_i=\rho_i\,\kappa_i\delta u_{mi}\,\partial u_i/\partial y$, where $\kappa_i$ is the coefficient of turbulent mixing for the *i*-th phase, $\delta$ is the width of the mixing layer. The polynomial approximations for the velocity profiles and other functions have been obtained based on the boundary conditions (2) [1, 2], e.g.

$$u_1/u_{01}=1-4\eta^3+3\eta^4,\quad u_2/u_{02}=1-6\eta^2+8\eta^3-3\eta^4, \qquad (3),(4)$$

$$B_1=B_1^{(0)}=1-\eta^3+0.5\eta^2(1-\eta)h(x),\qquad h\in[-6,0],$$
$$B_1=B_1^{(1)}=1-4\eta^3+3\eta^4+0.5\eta^2(1-\eta)^2h(x),\qquad h\in[-12,-6],$$
$$B_1=B_1^{(2)}=1-10\eta^3+15\eta^4-6\eta^5+0.5\eta^2(1-\eta)^3h(x),\qquad h\in[-20,-12],$$
$$B_1=B_1^{(3)}=1-20\eta^3+45\eta^4-36\eta^5+10\eta^6+0.5\eta^2(1-\eta)^4h(x),\qquad h\in[-30,-20], \qquad (5)$$
$$B_1=B_1^{(4)}=1-35\eta^3+105\eta^4-126\eta^5+70\eta^6-15\eta^7+0.5\eta^2(1-\eta)^5h(x),\qquad h\in[-42,-30],$$
$$B_1^{(5)}=1-56\eta^3+210\eta^4-336\eta^5+280\eta^6-120\eta^7+21\eta^8+0.5\eta^2(1-\eta)^6h(x),\qquad h\in[-56,-42],$$
$$B_1^{(6)}=1-84\eta^3+378\eta^4-756\eta^5+840\eta^6-540\eta^7+189\eta^8-28\eta^9+0.5\eta^2(1-\eta)^7h(x),\qquad h\in[-72,-56],$$

where $\eta=(y-y_0)/\delta$. Function $h(x)=\left(\partial^2 B_1/\partial\eta^2\right)_{\eta=0}$ is responsible for the variation of $B_1$ by $x$, it may vary in the range $h\le 0$ due to requirements of its nature.

The first approximation $B_1(\eta)$ in (5) reveals restricted application in the range $h\in[-6,0]$, while outside of this region it does not satisfy the boundary conditions (2) and the condition $0\le B_1(\eta)\le 1$, $\forall\eta\in[0,1]$ . Therefore, all the next approximations $B_1(\eta)$ in (5) were obtained as a transition of the piecewise continuous function-indicator $B_1^{(n)}$ to its next approximation following the condition that the derivative by $\eta$ with respect to a point $\eta=1$ is zero up to ($n$+1)-th order. These functions have the breaks at the transition points of the permanent characteristic function $B_1^{(n)}(\eta,h)$ from the one regional approximation to the other one (a first derivative has break at those points). The advantage of such approximations is that all functions $B_1^{(n)}$ are smoothly transforming from one region by *h(x)* to the next one as shown in Fig. 3. Each function $B_1^{(n)}$ exactly coincides with the previous one $B_1^{(n-1)}$ at the conjugation boundary, where the $B_1^{(n-1)}$ ends and the $B_1^{(n)}$ starts. Physical meaning of the varying approximations $B_1^{(n)}$ is determined by dependence of the phase distribution in the mixing layer on the density ration of the mixing phases: the higher is density of the surrounding liquid, the shorter is penetration of the first, lighter phase, into the mixing layer.

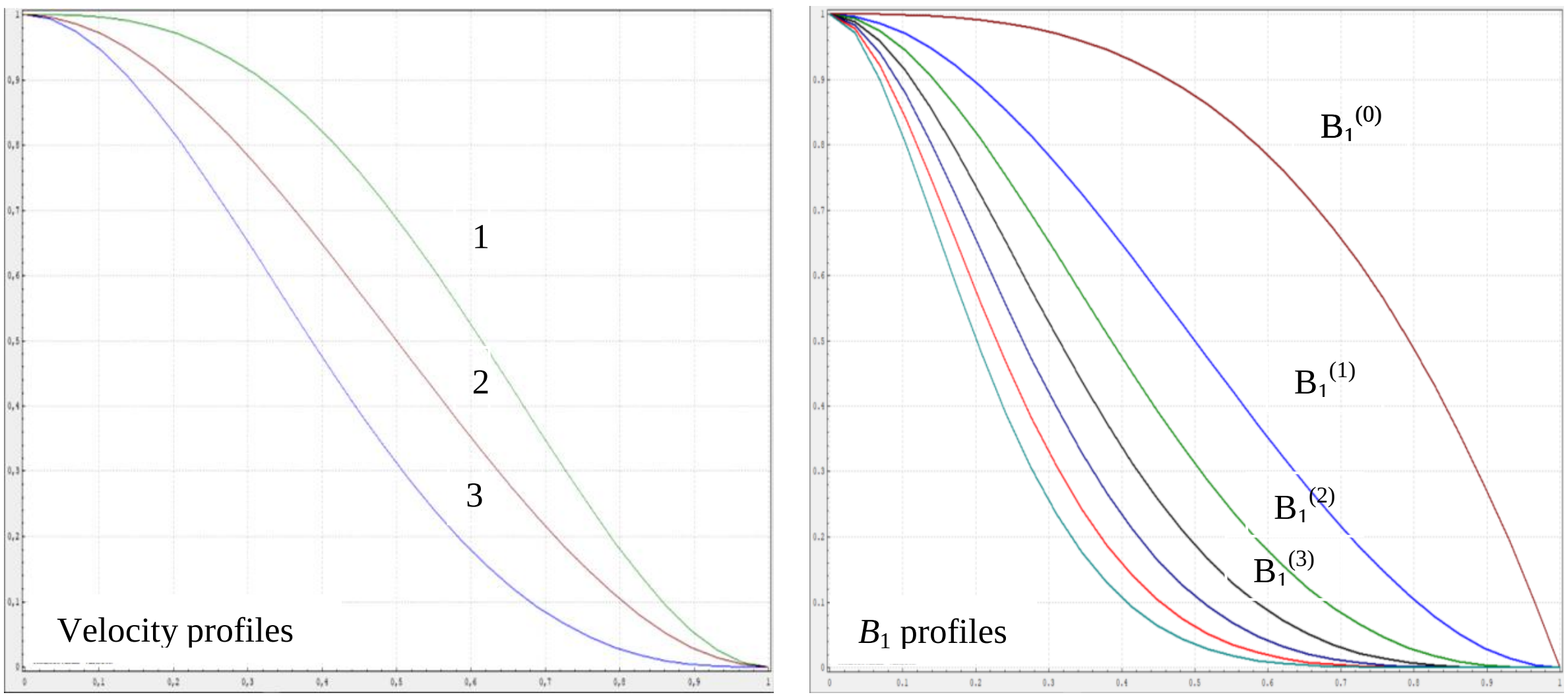

Fig. 3 Velocity and function-indicator $B_1^{(n)}$ profiles across the layer: 1-$u_1/u_{01}$, 2-$u_1/u_{m1}$, 3-$u_2/u_{02}$

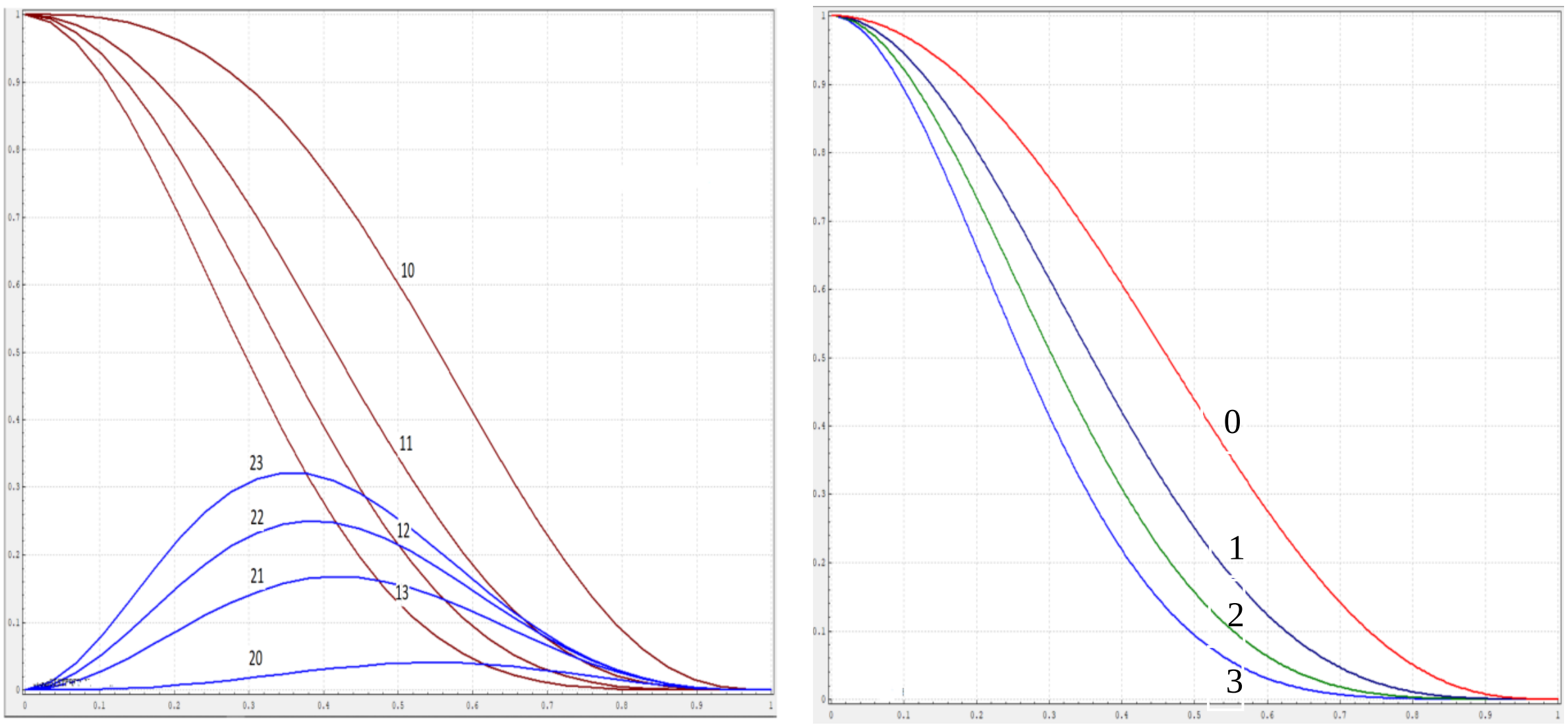

Fig. 4 The real velocity profiles of the phases for a few values of function $h(x)$: 0, -6, -12, -20 to the left 10-13: $B_1 u_1/u_{01}$, 20-23: $B_2 u_2/u_{02}$; to the right 0-3: $B_1 u_1/u_{m1}$

The disadvantage of the above approximations is a growing complexity of the functions $B_1^{(n)}$ in analytical calculation, which is nevertheless easily fought by modern computer analytical calculations. It is impossible to get a common approximation for $B_1(\eta, h)$ satisfying the boundary conditions (2) in all range by parameter $i_0$ (due to the requirement of variation $B_1$ from 0 to 1). The polynomial approximations for $u_2$, $B_1$ remain the same on a ground part of the jet but for the function $u_1$ the approximation is

$$u_1/u_{m1} = 1 - 3\eta^2 + 2\eta^3 . \qquad (6)$$

All polynomial approximations (3)-(6) are shown in Fig. 3. The real profiles for the multiphase flow are represented as a product of the corresponding profile multiplied by its function-indicator. The ones are presented in Fig. 4 for a few regions by the function $h(x)$.

## 3. INTEGRAL CORRELATIONS FOR THE INITIAL AND GROUND PARTS OF THE JET

Based on (3)-(5), the integral correlations have been derived for the two-phase turbulent jet [1, 2] integrating the mass and momentum conservation equations (1) with the boundary conditions (2) for the total cross-section of a flow $y=y_0+\delta$, as well as the momentum conservation for $y=y^*$, respectively:

$$u_{01}\left(r_0^2-y_0^2\right)=2\delta\int_0^1 B_1u_1\left(y_0+\delta\eta\right)d\eta\,,\quad \rho_1u_{01}^2\left(r_0^2-y_0^2\right)=2\delta\int_0^1\left(\rho_1B_1u_1^2+\rho_2B_2u_2^2\right)\left(y_0+\delta\eta\right)d\eta\,,$$

$$\rho_1u_{01}\left(u_{01}-u_1^*\right)y_0y_0'+\frac{d}{dx}\delta\int_0^{\eta^*}\sum_{j=1}^{2}\rho_jB_ju_j^2\left(y_0+\delta\eta\right)d\eta-\sum_{j=1}^{2}u_i^*\frac{d}{dx}\delta\int_0^{\eta^*}\rho_iB_iu_i\left(y_0+\delta\eta\right)d\eta=$$

$$=\left(y_0+\delta\eta^*\right)\sum_{j=1}^{2}\rho_jB_j\kappa_ju_{0j}\frac{\partial u_j^*}{\partial\eta},\qquad B_1+B_2=1\,. \tag{7}$$

The first equation in (7) was got integrating the mass conservation equation by $y$, the second and the third ones – integrating the momentum conservation for the total flow of a two-phase mixture for $y=y_0+\delta$ and $y=y^*$, respectively. The parameters at $\eta=\eta^*<1$ are signed with a star *.

The integral correlations for the ground part of a jet are [1]:

$$2\int_0^{\delta}B_1u_1ydy=u_{01}r_0^2\,,\qquad 2\sum_{j=1}^{2}\int_0^{\delta}\rho_jB_ju_j^2ydy=\rho_1u_{01}^2r_0^2\,, \tag{8}$$

$$\frac{d}{dx}\sum_{j=1}^{2}\int_0^{y^*}\rho_jB_ju_j^2ydy-\sum_{j=1}^{2}u_j^*\frac{d}{dx}\int_0^{y^*}\rho_jB_ju_jydy=y^*\sum_{j=1}^{2}B_j^*\tau_j^*\,,$$

where the first is equation of mass conservation for the first phase, the second and the third – the momentum conservation equations for the total and for the part of the cross section, respectively, according to the methodology [2].

The integral correlations for the ground part of a jet were obtained similarly to the initial part. The momentum equations for the total and for the part of the cross section, respectively, were got according to [11]. The momentum equation on the jet's axis ($y$=0) is used too:

$$\sum_{j=1}^{2}\rho_jB_{mj}u_{mj}\frac{du_{mj}}{dx}=2\sum_{j=1}^{2}\left[\frac{\partial}{\partial y}\left(B_j\tau_j\right)\right]_m\,. \tag{9}$$

Mathematical model (7)-(9) including the ordinary differential equations by longitudinal coordinate $x$ are used for analysis and numerical simulation of the basic features of turbulent stationary two-phase jet of two immiscible liquids. The function-indicator $B_1$ shows how much is a presence of the first phase in a selected point of mixing zone, which can be directly compared to an experimental data by two-phase sensor [1]. Therefore, a solution of the task may give both parameters of the flow together with their belonging to a particular phase.

## 4. TRANSFORMATION OF THE MATHEMATICAL MODEL TO DIMENSIONLESS FORM

The equation array (7) for the initial part of the jet is transformed to the following dimensionless form with the scales $r_0$, $\delta$, $u_{0i}$ for the longitudinal and transversal coordinates and velocity, respectively:

$$y_0^2+2\delta\sum_{j=1}^{2}y_0^{2-j}\delta^{j-1}a_j=1\,,\qquad y_0^2+2\delta\sum_{j=1}^{2}y_0^{2-j}\delta^{j-1}\left(a_{j+2}+i_0b_{j+2}\right)=1\,, \tag{10}$$

$$\left(1-u_1^*\right)y_0\frac{dy_0}{d\varsigma}+\frac{d}{d\varsigma}\delta\sum_{j=1}^{2}y_0^{2-j}\delta^{j-1}\left(a_{j+2}^*+i_0b_{j+2}^*\right)-\frac{d}{d\varsigma}\delta\sum_{j=1}^{2}y_0^{2-j}\delta^{j-1}\left(a_j^*u_1^*+i_0b_j^*u_2^*\right)=$$

$$=\left(y_0+\delta\eta^*\right)\sum_{j=1}^{2}B_j^*\left(\frac{\partial u_j}{\partial\eta}\right)^*\left(i_0\kappa_{21}\right)^{j-1}.$$

Here are the star marked values are taken by $\eta=\eta^*$. Normally it is adopted $\eta^*=0.5$. The other assignment are as follows:

$$\bar{y}_0=y_0/r_0,\ \bar{\delta}=\delta/r_0,\ \eta=\left(y-y_0\right)/\delta,\ \bar{x}=x/r_0,\ \varsigma=\kappa_1\bar{x},\ s_0=u_{02}/u_{01},\ i_0=ns_0^2,$$

$$n=\rho_2/\rho_1,\ \kappa_{21}=\kappa_2/\kappa_1,\ a_i=a_{i1}+a_{i2}h,\ b_i=b_{i1}+b_{i2}h, \qquad (11)$$

$$a_i=\int_0^1 B_1\bar{u}_1\eta^{j-1}d\eta,\ b_i=\int_0^1 B_2\bar{u}_2\eta^{j-1}d\eta\ (i=1,2);\ a_i=\int_0^1 B_1\bar{u}_1^2\eta^{j-1}d\eta,\ b_i=\int_0^1 B_2\bar{u}_2^2\eta^{j-1}d\eta\ (i=3,4); j=1,2.$$

The computed values of the integral parameters in (11) have been done in a range of variation of the function $h(x)$ according to the approximation of the velocity profiles (3), (4) and the function-indicator of the phase $B_1$ (5). As shown below mostly the region by parameter $h(x)$ is covered in (5) for the bright enough density ratio of the mixing liquid phases. The coefficients $a_{ij}$, $b_{ij}$ according to (11), (3)-(5) are presented in the Table 1 and Table 2:

Table 1 – Integral parameters $a_{ij}$ of the model for different regions of the function-indicator $B_1$

| Value $a_{ij}$ for $h$: | $a_{11}$ | $a_{12}$ | $a_{21}$ | $a_{22}$ | $a_{31}$ | $a_{32}$ | $a_{41}$ | $a_{42}$ |
|---|---|---|---|---|---|---|---|---|
| $h\in[0,-6]$ | 0.5464 | 0.0208 | 0.1667 | 0.0101 | 0.4604 | 0.0139 | 0.1198 | 0.0059 |
| $[-6,-12]$ | 0.4857 | 0.0107 | 0.1333 | 0.0046 | 0.4250 | 0.0080 | 0.1027 | 0.0030 |
| $[-12,-20]$ | 0.4310 | 0.0062 | 0.1065 | 0.0023 | 0.3884 | 0.0049 | 0.0866 | 0.0017 |
| $[-20,-30]$ | 0.3844 | 0.0038 | 0.0859 | 0.0013 | 0.3542 | 0.0032 | 0.0728 | 0.0010 |
| $[-30,-42]$ | 0.3455 | 0.0025 | 0.0703 | 0.00078 | 0.3237 | 0.0022 | 0.0614 | 0.00064 |
| $[-42,-56]$ | 0.3128 | 0.0017 | 0.0583 | 0.00049 | 0.2969 | 0.0016 | 0.0522 | 0.00042 |
| $[-56,-72]$ | 0.2853 | 0.0013 | 0.0490 | 0.00032 | 0.2734 | 0.0012 | 0.0447 | 0.00028 |

Table 2 – Integral parameters $b_{ij}$ of the model for different regions of the function-indicator $B_1$

| Value $b_{ij}$ for $h$: | $b_{11}$ | $b_{12}$ | $b_{21}$ | $b_{22}$ | $b_{31}$ | $b_{32}$ | $b_{41}$ | $b_{42}$ |
|---|---|---|---|---|---|---|---|---|
| $h\in[0,-6]$ | 0.0179 | -0.0101 | 0.0095 | -0.0042 | 0.0054 | -0.0048 | 0.0023 | -0.0016 |
| $[-6,-12]$ | 0.0429 | -0.0060 | 0.0214 | -0.0022 | 0.0149 | -0.0032 | 0.0059 | -0.00097 |
| $[-12,-20]$ | 0.0690 | -0.0038 | 0.0327 | -0.0012 | 0.0625 | -0.0023 | 0.0098 | -0.00062 |
| $[-20,-30]$ | 0.0939 | -0.0025 | 0.0424 | -0.00076 | 0.0390 | -0.0016 | 0.0141 | -0.00042 |
| $[-30,-42]$ | 0.1167 | -0.0018 | 0.0506 | -0.00049 | 0.0515 | -0.0012 | 0.0180 | -0.00029 |
| $[-42,-56]$ | 0.1371 | -0.0013 | 0.0573 | -0.00032 | 0.0637 | -0.00092 | 0.0215 | -0.00021 |
| $[-56,-72]$ | 0.1552 | -0.00096 | 0.0629 | -0.00022 | 0.0752 | -0.00072 | 0.0246 | -0.00015 |

The integral correlation for the part of the mixing layer in the system (10) was considered at $\eta=0.5$. The corresponding coefficients $a_{ij}^*$, $b_{ij}^*$ for this middle section are given in the Tables 3, 4:

Table 3 – Integral parameters $a_{ij}^*$ of the model for different regions of the function-indicator $B_1$

| Value $a_{ij}^*$ for $h$: | $a_{11}^*$ | $a_{12}^*$ | $a_{21}^*$ | $a_{22}^*$ | $a_{31}^*$ | $a_{32}^*$ | $a_{41}^*$ | $a_{42}^*$ |
|---|---|---|---|---|---|---|---|---|
| $h \in [0,-6]$ | 0.4436 | 0.0110 | 0.1029 | 0.0038 | 0.4103 | 0.0094 | 0.0902 | 0.0032 |
| $[-6,-12]$ | 0.4206 | 0.0071 | 0.0941 | 0.0024 | 0.3912 | 0.0062 | 0.0831 | 0.0020 |
| $[-12,-20]$ | 0.3920 | 0.0062 | 0.0836 | 0.0015 | 0.3673 | 0.0042 | 0.0745 | 0.0012 |
| $[-20,-30]$ | 0.3619 | 0.0033 | 0.0730 | 0.00097 | 0.3416 | 0.0029 | 0.0657 | 0.00084 |
| $[-30,-42]$ | 0.3327 | 0.0023 | 0.0631 | 0.00064 | 0.3164 | 0.0021 | 0.0574 | 0.00056 |
| $[-42,-56]$ | 0.3069 | 0.0016 | 0.0548 | 0.00043 | 0.2936 | 0.0015 | 0.0503 | 0.00039 |
| $[-56,-72]$ | 0.2814 | 0.0012 | 0.0468 | 0.00030 | 0.2711 | 0.0011 | 0.0434 | 0.00027 |

Table 4 – Integral parameters $b_{ij}^*$ of the model for different regions of the function-indicator $B_1$

| Value $b_{ij}^*$ for $h$: | $b_{11}^*$ | $b_{12}^*$ | $b_{21}^*$ | $b_{22}^*$ | $b_{31}^*$ | $b_{32}^*$ | $b_{41}^*$ | $b_{42}^*$ |
|---|---|---|---|---|---|---|---|---|
| $h \in [0,-6]$ | 0.0075 | -0.0142 | 0.0028 | -0.0047 | 0.0039 | -0.0085 | 0.0014 | -0.0025 |
| $[-6,-12]$ | 0.0214 | -0.0048 | 0.0079 | -0.0015 | 0.0114 | -0.0030 | 0.0039 | -0.00083 |
| $[-12,-20]$ | 0.0392 | -0.0033 | 0.0141 | -0.00096 | 0.0214 | -0.0022 | 0.0071 | -0.00057 |
| $[-20,-30]$ | 0.0585 | -0.0023 | 0.0206 | -0.00064 | 0.0328 | -0.0016 | 0.0106 | -0.00039 |
| $[-30,-42]$ | 0.0777 | -0.0017 | 0.0267 | -0.00044 | 0.0446 | -0.0012 | 0.0140 | -0.00028 |
| $[-42,-56]$ | 0.0954 | -0.0012 | 0.0320 | -0.00030 | 0.0560 | -0.00091 | 0.0172 | -0.00020 |
| $[-56,-72]$ | 0.1130 | -0.00094 | 0.0372 | -0.00022 | 0.0675 | -0.00071 | 0.0202 | -0.00015 |

Except the above, for the dimensionless parameters we retain the same notations as for the dimensional ones. Only here in (11) it is stated for clarification of the dimensionless notations.
The sliding factor $s_0$ is supposed to be constant. Boundary conditions (2) for (11) are transformed as

$$\zeta=0,\ \ y_0=1,\ \ \delta=0; \qquad \zeta=\zeta_i,\ \ y_0=0,\ \ \delta=\delta_i; \tag{12}$$

where $\zeta_i$, $\delta_i$ are the dimensionless length of a jet and its maximal radius (at the end of the initial part).

The characteristics $y_0(\zeta)$, $\delta(\zeta)$, $h(\zeta)$ are computed from (10), (12). Then all other are got for the stated values of the main parameters: $i_0$, $\kappa_1$, $\kappa_2$. The first parameter is slightly indefinite due to difficulties with estimation of the phases' sliding, while the other two are known from the experimental studies for the specific conditions. The main problem with validation of the mathematical model against the experimental data is correct estimation of the coefficients of turbulent mixing $\kappa_1$, $\kappa_2$ in each specific case. But the advantage of the model is a possibility to have all characteristics of a flow together with their belonging to a particular phase through the functions $B_1$, $B_2$. Distribution of the transversal velocities, interface interactions, coefficients of the volumetric $q$ and mass ejection $g$, and kinetic energy $e_i$ for the phases in a flow are computed as follows [1]:

$$\frac{B_1 v_1}{\kappa_1 u_{01}} = B_1 u_1 \frac{d}{d\varsigma}\left(y_0+\delta\eta\right) - \frac{1}{y_0+\delta\eta}\frac{d}{d\varsigma}\left(0.5y_0^2 + a_1 y_0\delta + a_2\delta^2\right),$$

$$\frac{B_2 v_2}{\kappa_1 u_{02}} = B_2 u_2 \frac{d}{d\varsigma}\left(y_0+\delta\eta\right) - \frac{1}{y_0+\delta\eta}\frac{d}{d\varsigma}\left(b_1 y_0\delta + b_2\delta^2\right), \tag{13}$$

$$q = 2s_0\delta\left(b_1 y_0 + b_2\delta\right),\ \ e_1 = y_0^2 + 2\int_0^1 B_1\bar{u}_1^3\left(y_0+\delta\eta\right)\delta d\eta,\ \ e_2 = 2i_0 s_0\int_0^1 B_2\bar{u}_2^3\left(y_0+\delta\eta\right)\delta d\eta,$$

$$R_{21}=-\frac{1}{\left(y_0+\delta\eta\right)\delta}\frac{\partial u_1}{\partial\eta}\left\{\frac{d}{d\varsigma}\left(0.5y_0^2+a_1y_0\delta+a_2\delta^2\right)+\frac{\partial}{\partial\eta}\left[\left(y_0+\delta\eta\right)B_1\frac{\partial u_1}{\partial\eta}\right]\right\}.$$

For the short transient part of the jet there is no developed substantiated scheme, therefore it is not under consideration here. The ground part is considered after a short transient part. The method is well elaborated and supported with the experimental data [1,2]. The dimensionless equation array (8), (9) for the ground part is the next

$$2B_{m1}u_{m1}\delta^2\sum_{j=1}^{2}\alpha_{1j}h^{j-1}=1,\quad 2B_{m1}u_{m1}\delta^2\sum_{j=1}^{2}\left(\alpha_{2j}+i_0\beta_{2j}\right)h^{j-1}+i_0\beta_{20}=1\,, \tag{14}$$

$$\frac{d}{d\varsigma}u_{m1}^2\delta^2\left[B_{m1}\sum_{j=1}^{2}\left(\alpha_{2j}^*+i_0\beta_{2j}^*\right)h^{j-1}+i_0\beta_{20}^*\right]-u_{m1}\frac{d}{d\varsigma}\left(u_{m1}\delta^2\right)\left[B_{m1}\sum_{j=1}^{2}\left(\alpha_{1j}^*u_1^*+i_0\beta_{1j}^*u_2^*\right)h^{j-1}+i_0u_2^*\beta_{10}^*\right]=$$

$$=\eta^*\delta u_{m1}^2\left[\left(1-i_0\kappa_{21}\right)B_{m1}\sum_{j=1}^{2}\left(\frac{\partial u_j}{\partial\eta}\right)^*\gamma_j^*h^{j-1}+i_0\kappa_{21}\left(\frac{\partial u_2}{\partial\eta}\right)^*\right].$$

Here are:

$$\bar{x}=\frac{x-x_t}{r_0},\ \bar{u}_{mi}=\frac{u_{mi}}{u_{0i}},\ \bar{u}_i=\frac{u_i}{u_{mi}},\ i_0=ns_0^2\,,\ \bar{B}_2=\frac{B_2}{B_{m1}},\ \bar{B}_1=\frac{B_1}{B_{m1}}=\gamma_1+\gamma_2h,$$

$$\int_0^1\bar{B}_1\bar{u}_1^i\eta d\eta=\sum_{j=1}^{2}\alpha_{ij}h^{j-1}\,,\ \int_0^1\bar{B}_2\bar{u}_2^i\eta d\eta=\frac{\beta_{i0}}{B_{m1}}+\beta_{i1}+\beta_{i2}h\quad(i=1,\,2). \tag{15}$$

The same as previously, we use these notations for dimensionless values only here and keep previous assignments for the dimensionless parameters as for the dimension ones in all other equations. Star * means a value by $\eta=\eta^*<1$, $x_t$ is the length of a transient part of a jet flow. It is assumed $u_{m2}=s_0u_{m1}$ ($s_0$=const), which means that sliding of the phases is preserved the same as for the initial part of a jet.

The boundary conditions for the equation array (14) are stated in a form

$$\zeta=0,\ u_{m1}=1,\ B_{m1}=1,\ \delta=\delta_t;\qquad \zeta=\infty,\ u_{m1}=0,\ B_{m1}=0,\ \delta=\infty; \tag{16}$$

$\delta_t$ is a radius of the jet at the transient cross section. Solution of the boundary problem (14), (16) yields the functions $u_{m1}(\zeta)$, $B_{m1}(\zeta)$, $\delta(\zeta)$ and $h(\zeta)$ for the stated values $i_0$, $\kappa_{21}$. Then all parameters of a two-phase flow are got: turbulent stresses, mass flow rate, trajectories of the phases in a mixing layer.

## 5. CONCLUSIONS AND FURTHER RESEARCH OF THE TURBULENT TWO-PHASE JETS

The conducted study of the mixing processes in a turbulent heterogeneous jet of immiscible liquids allows determining the main parameters of the devices including distribution of the parameters along the axis of the devices, as well as across the mixing layer. This is important for the optimal organization of the working process. The results may be useful in a number of chemical technology and other engineering fields, where jet hydraulic machines are applied. The mathematical model developed for the free and confined jets of two-phase flows and the approximate correlations proposed from the analysis are available for implementation into the research and engineering calculations by scientists and engineers. Comparison with experiments was done for earlier for a few different real situations, e.g. example [13, 20] is given in Fig. 5.

The problem of turbulent mixing in jet machines working on mutually immiscible liquids is quite complex and not so good investigated. Therefore, this attempt can be interesting both, in theoretical and practical aspects because it revealed some basic features of the multiphase flow. Later on, it will be continued to understand more in deep the processes, some regularities of which were already understood on the mathematical models developed and numerical computer simulations performed. The earlier obtained results are enforced now and substantiated more in deep including expanding of the range of the method applicability. The new numerical models were developed and the next paper will outline the results obtained.

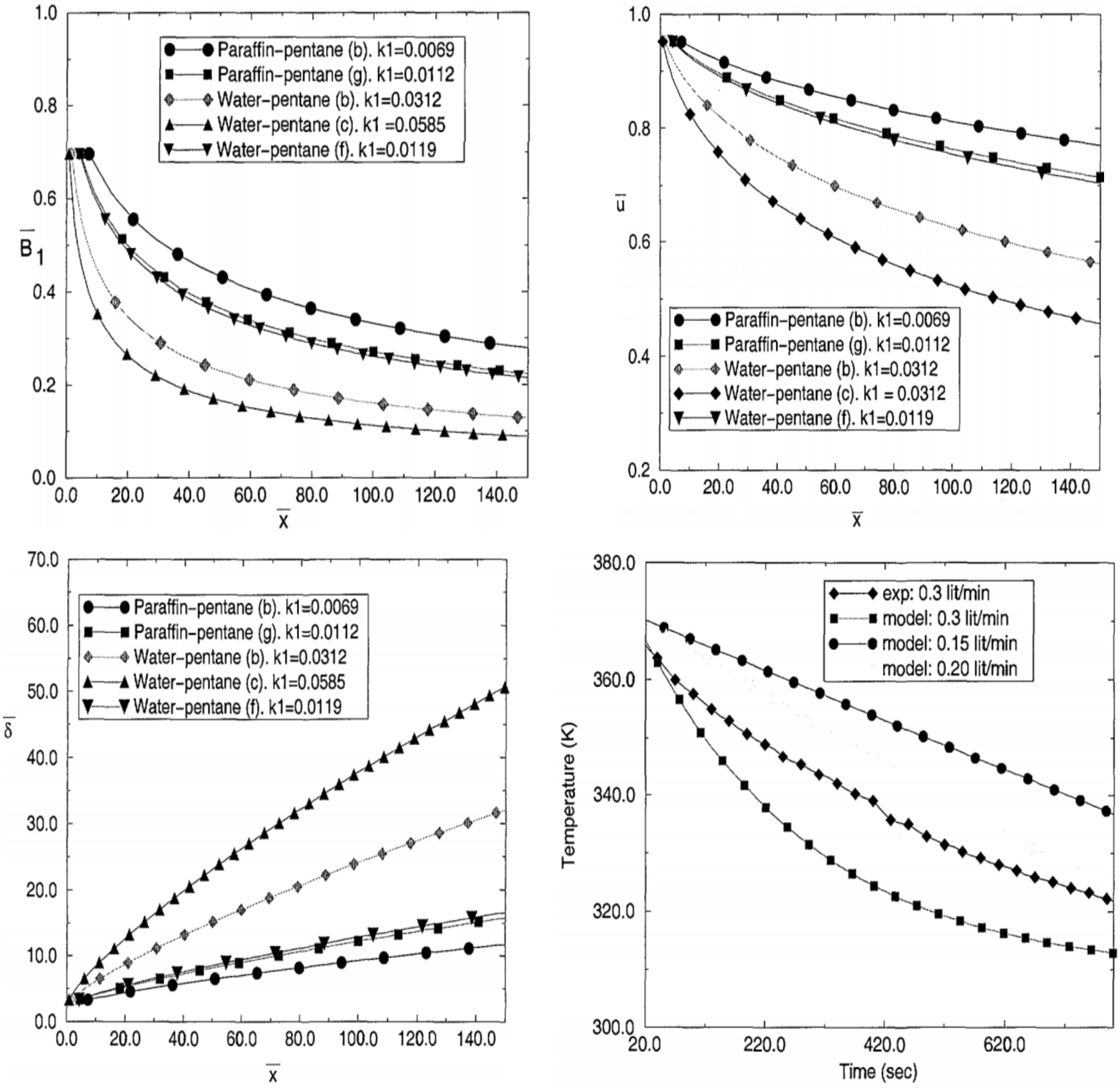

Fig. 5 Function-indicator $B_1$, flow velocity, width of mixing zone and temperature against experiment

**Acknowledgements**

The author devotes this article to the blessed memory of Professor A.I. Nakorchevskii, who developed the method for modeling the multiphase turbulent jets. And the author wishes to acknowledge Professor Torsten H. Fransson for possibility to work at the Dept of Energy Technology (KTH), during many years since 2001.